\documentclass[twocolumn,showpacs,preprintnumbers,amsmath,amssymb]{revtex4}
\usepackage{graphicx}
\usepackage{amsmath}
\usepackage{amssymb}

\begin{document}

\title{Meaning of the splitting process for the transition to self-sustained turbulence in a magnetized cylindrical plasma}

\author{S. Knauer$^1$, C. Moon$^{2,3}$, T. Schleyerbach$^1$, S. Inagaki$^4$, N. Fahrenkamp$^1$, A. Fujisawa$^{2,3}$, P. Manz$^1$}
\affiliation{$^1$ Institute of Physics, University of Greifswald, Felix-Hausdorff-Str.6, 17489 Greifswald, Germany \\
$^2$ Research Center for Plasma Turbulence, Kyushu University, Kasuga 816-8580, Japan \\
$^3$ Research Institution for Applied Mechanics, Kyushu University, Kasuga 816-8580, Japan \\
$^4$ Institute of Advanced Energy, Kyoto University, Gokasho, Uji, Kyoto 611-001, Japan
} 

\date{\today}

\begin{abstract}
When turbulent structures split more frequently before they decay, persistent turbulence forms in neutral fluid shear flows. Whether this concept can be extended to linear magnetized plasmas is investigated here and compared to the behavior of the pipe flow. With increasing control parameter the dynamics in the magnetized plasmas is known to undergo several changes from a quasiperiodic to a phase locked to a weakly turbulent regime. When the phase-locked regime breaks down, the splitting time approaches the decreasing lifetime reflecting self-sustained turbulence, as known from the pipe flow.
\end{abstract}

\maketitle

For the transition into turbulence two scenarios are distinguished, nowadays. The first kind of transition to turbulence occurs through linear instabilities successively growing in size and number, with the flow becoming more and more disordered in space and time. With the aid of a modal theory, a critical Reynolds number can be predicted, at which the linear growth rates of an instability get positive. If the onset of turbulence matches this critical Reynolds number, this kind of transition is called super-critical transition. Examples for flows showing such a super-critical transition to turbulence are the Rayleigh-Bénard and the Taylor-Couette flow.

The second kind of transition to turbulence is enabled by finite amplitude disturbances, which can sustain themselves. Because of the amplitude dependence, the transition here is rather abrupt. In some cases there might be no linear instability at all. It can appear for Reynolds numbers much lower than the linear instability threshold, and is, thus, referred to as sub-critical transition to turbulence. The transition to turbulence in the pipe flow is one particular prominent example.
The transition from laminar to turbulent flows in pipe flows was already studied more than 100 years ago by Reynolds \cite{REYNOLDS1883}. He already observed that at low Reynolds numbers turbulence occur in spatially localized finite amplitude disturbances. In the context of pipe flows these are called puffs. These puffs are metastable and decay with a characteristic time scale $\tau_l$ \cite{AVILA2010}. From transient to persistent turbulence the lifetime does not diverge \cite{HOF2006} as assumed before. Persistence is achieved through a different mechanism. A puff can split and generate new puffs. Also puff splitting exhibits a characteristic time scale $\tau_{sp}$ \cite{AVILA2011}. The point where these time scales are equal is referred to persistent or self-sustaining turbulence.

We study turbulence in magnetized plasmas. Turbulent transport is responsible for the main part of the particle and energy losses and is therefore an important research topic in magnetically confined fusion research. In large confinement experiments as tokamaks or stellarators, already in the start up phase of the discharge, steep gradients appear because of the confinement. In presence of magnetic curvature these gradients lead directly to violent instabilities of the interchange type. The quasi-stationary equilibrium is determined by the balance of particle and heat sources and the particle and heat transport. Since the transport is dominated by turbulence, the plasma is in a turbulent state under quasi-stationary equilibrium conditions. Fully developed broadband turbulence is usually observed and transition to turbulence studies are very difficult. 

The transition to a turbulent state can be studied in laboratory-scale plasma experiments. Here linear instabilities such as drift-waves can studied  \cite{BROCHARD2005,YAMADA2008} which are highly relevant for turbulence in large confinement experiments. Drift wave turbulence is the simplest case of turbulence in magnetized plasmas \cite{HASEGAWA83,WAKATANI84}. Evidence of the Landau-Ruelle-Takens scenario is found for drift-waves \cite{KLINGER97,KLINGER97a,GRULKE2002,MANZ2011a} and flute instabilities \cite{BROCHARD2006a} in cylindrical magnetized plasmas. The Landau-Ruelle-Takens scenario is usually associated with the super-critical transition.
However, drift wave turbulence can be self-sustaining \cite{SCOTT1992,MANZ2011a,MANZ2018}, which is the main property of sub-critical turbulence. Here we want to investigate if we can find signs of puff-splitting-like behavior as observed in the pipe flow in cylindrical magnetized plasmas at the onset of turbulence. 
 
Experiments have been carried out in the linear magnetized plasma experiment PANTA \cite{YAMADA2008} of a length of $4.05\,\mathrm{m}$ and a diameter of $45\,\mathrm{cm}$. A homogeneous axial magnetic field is generated by a set of Helmholtz coils. The plasma is heated by a double loop helicon source with a diameter of $10$ cm at one end of the device. The typical cylindrical plasma radius is $5\,\mathrm{cm}$. At the other end the plasma terminates at a stainless steel end-plate. 

For the pipe flow the Reynolds number is the control parameter. But the viscosity in plasmas is negligibly low, thus, the Reynolds number is not relevant here. Drift-waves are driven unstable by electron-ion collisions and a normalized collisionality $C$ can be used as the control parameter for drift-wave turbulence \cite{HASEGAWA83,WAKATANI84} as for example used in Refs.~\cite{MANZ2008,SCHMID2017,GARLAND2017}. However, the collisionality $C$ takes over two properties, driving and damping, in contrast to the Reynolds number, and it is therefore not a good parameter to replace it. A third approach, which is used in particular in helicon heated plasmas is to utilize the magnetic field strength $B$ \cite{BURIN2005,MANZ2011a,THAKUR2014}. In helicon heated discharges like in PANTA the drift-wave is driven by the density gradient. At the radial boundaries there is no plasma.  The density gradient between the plasma and vacuum raises with increasing magnetic field strength and, thus, leads to enhanced drive of the drift-wave. It is also the limited inertial range that makes the development of broadband drift-wave turbulence impossible. The size of the drift-wave decreases with increasing magnetic field strength \cite{RAMISCH2005a}. No large drift-waves fit into the experiment at low magnetic fields. To study the route to turbulence, a magnetic field scan from $B=0.05$ to $0.13\,\mathrm{T}$ is done at a heating power of $3\,\mathrm{kW}$ and $7\,\mathrm{MHz}$ radio frequency at an Argon pressure of $0.53\,\mathrm{Pa}$. 

There are important differences to pipe flows to be made. In the pipe, the puffs are advected downstream in the axial direction. And although, in the cylindrical plasma experiment, the plasma is also advected downstream in the axial direction along the magnetic field line, the turbulence is perpendicular to the magnetic field. In other words, the turbulence-structures in the plane perpendicular to the magnetic field are mainly advected in azimuthal direction.
Therefore, the dynamics are not investigated in the axial direction, as one would expect from the pipe, but in the azimuthal direction.
Furthermore, in the pipe flow experiments reported in Ref.~\cite{AVILA2011}, perturbations are injected in a controlled manner into an otherwise very quiescent flow. This is not the case in the plasma column, where the initial conditions are not in a quiescent state. 

Fluctuations are measured by a 64-channel Langmuir probe array \cite{YAMADA2007a} which is installed at an axial distance of $z=2.125\,\mathrm{m}$ away from the source. The Langmuir probes made of tungsten wire with a diameter of $0.8\,\mathrm{mm}$ and length of $3\,\mathrm{mm}$ are aligned on a circle of radius $r=4\,\mathrm{cm}$. The azimuthal distance between the probes is $3.9\,\mathrm{mm}$. The probe tips measured alternating ion saturation current and floating potential fluctuations, which are usually considered to be a good approximation for plasma density and potential fluctuations, respectively. Here, only the data from the floating potential is shown. The distance $\Delta y$ between two floating tips is $7.8\,\mathrm{mm}$. 


\begin{figure}[tbh]
   \begin{center}
	 \includegraphics[width=0.49\textwidth]{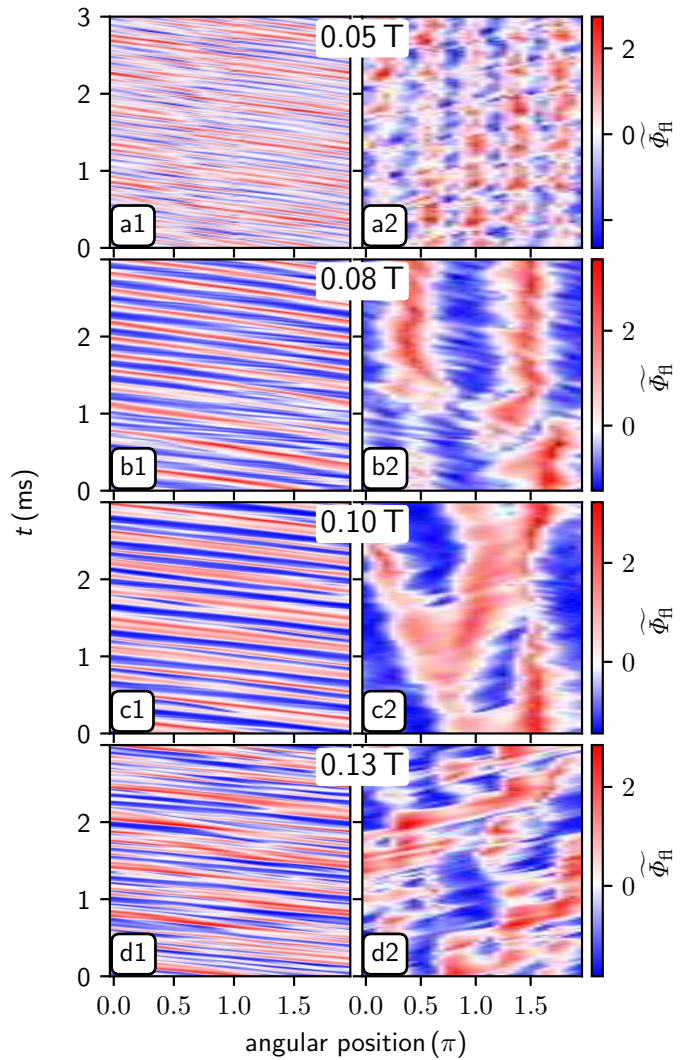}
   \caption{Space–time diagrams in the laboratory reference frame (l.h.s., number 1) and reference frame co-moving with the dominant mode (r.h.s., number 2) at different magnetic field strengths.} 
   \label{Fig_examples} 
   \end{center}
   \end{figure}

Fig.~\ref{Fig_examples} shows the floating potential fluctuations $\widetilde{\mathit{\Phi}_{\mathrm{fl}}}~=~(\mathit{\Phi}_{\mathrm{fl}} - \overline{\mathit{\Phi}_{\mathrm{fl}}} ) \, / \, \mathit{\sigma}_{\mathrm{fl}}$ (mean-free and normalized to its standard deviation), detected with the 64 channel probe array. The ordinate shows the time, the angular position is shown on the abscissa and the intensity of $\widetilde{\mathit{\Phi}_{\mathrm{fl}}}$ is color coded. In the left column (labeled with the number 1) the $\widetilde{\mathit{\Phi}_{\mathrm{fl}}}$ without further correction is presented. Because of the $E\times B$ background velocity and their phase velocity, the structures propagate in azimuthal direction, thus, unlike in the tube, the dynamic here is periodic. When a structure ends at the left side and reappears on the right side, it actually is the same puff. 

One can clearly see that the structures are quite long-living. Although often used precisely for this, the auto-correlation time $\tau_{ac}$ is not suitable for the determination of the lifetime of the structures. At a fixed location, $\tau_{ac}$ is largely determined by the background and phase velocity and the size of the structure and not by the lifetime of the perturbation. To calculate the lifetime of the perturbations they must be observed in a reference frame that moves along with it.

The right column of fig.~1 (labeled with the number 2) shows the data from the left column moved into an azimuthally co-moving frame. To achieve this, first the cross-correlation between neighboring probes probe tips is calculated.
We determine the time lag $\tau_{cc}$ by the maximum value of the cross-correlation. This provides the angular propagation speed $V=\Delta_y /\tau_{cc}$ with $\Delta y$ being the distance between the probe tips.   
The data is then averaged over the time lag $\tau_{cc}$. Next, the data with a time resolution of $\Delta t = \tau_{cc}$ is shifted for each time step in the azimuthal direction by one probe tip $\Delta y$. This corresponds to the transformation  $y^\prime = y - \Delta y $ equal to a transformation to the co-moving frame $y^\prime = y - \frac{\Delta y}{\Delta t} \Delta t $ with  the propagation speed $V=\Delta_y /\Delta t=\Delta_y /\tau_{cc}$. 

In the co-moving frame, at $B = 0.05\,\mathrm{T}$ the dynamics is dominated by waves, here with an azimuthal mode number of $m=5$. A pure $m=5$ wave would consist of five maxima constant in time. Here the amplitude is modulated in time and space. Nonlinear effects are clearly visible, but the linear properties come through more. This regime is called the quasi-linear regime in the Landau-Ruelle-Takens scenario. 

Surprisingly, with increasing control parameters ($B = 0.08\,\mathrm{T}$), the dynamics becomes first more coherent. The appearance of this long-lived structure is in good agreement to previous studies related to the Landau-Ruelle-Takens \cite{KLINGER97,MANZ2011a} scenario. In the so-called mode-locked regime, a quasi-coherent mode appears \cite{BURIN2005,MANZ2011a}. The quasi-coherent mode is a nonlinear mode. It is characterized by a high spatiotemporal symmetry, it appears particularly stable and long-living, also already known at PANTA as the solitary wave \cite{KOBAYASHI2017}. The quasi-coherent mode is destabilized by the occasional occurrence of phase defects \cite{KLINGER97,GRULKE2002}. These defects have a relatively short lifetime of the order of the period duration of the dominant mode and lead to the broadening of the spectrum around the dominant peaks \cite{KLINGER97}. When the phase defects occur more often, the spatiotemporal symmetry gets lost and the turbulent state develops. The individual regimes (quasi-linear, phase-locked and weak turbulent) are not strictly separated from each other. It has been found in Ref.~\cite{MANZ2011a} that the transitions to the different regimes are not strictly distinct. Features of one regime can be found before or after the other. Comparing the right and left column, we see that the phase defects correspond to the splits.

\begin{figure}[tbh]
    \centering
    \includegraphics[width=0.49\textwidth]{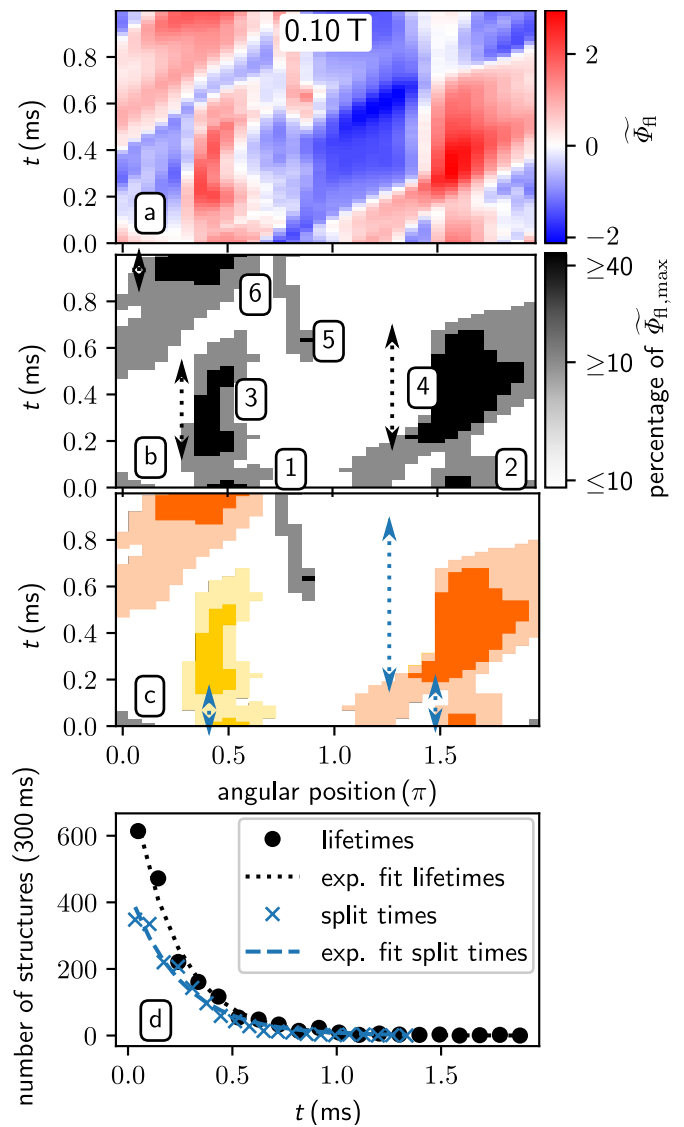}
	   \caption{Cohesive structures are identified by means of a flood-fill algorithm. Fig.~a shows $1\,\mathrm{ms}$ of $\widetilde{\mathit{\Phi}_{\mathrm{fl}}}$ in the co-moving frame as introduced in fig.~1 at $0.1\,\mathrm{T}$. Fig.~b translates the intensity into distinctive regimes. Structures are shown in black and are numbered with 1-6. Three lifetimes are indicated by double headed arrows. The paths between are shown in grey. Fig.~c illustrates the assignment of splits by giving dependent structures the same color. The resulting split times are indicated by double headed blue arrows. Fig.~d shows the binned life- and split times for a $300\,\mathrm{ms}$ time window and their respective exponential fit. For further information please see the text.} 
       \label{Fig_examples2} 
\end{figure}

At $B = 0.08\,\mathrm{T}$ the dynamics looks similar to the dynamics of puffs in a pipe flow. There are long lasting patches of enhanced potential perturbations. These patches occasionally split up. A fine structure connecting these two patches is also recognizable. 
Such fine connections can also be seen in the pipe flow \cite{AVILA2011}.  In the picture shown, it looks as if the left patch is generated by the right one. Puff-split-like events can be observed. The increase of splitting rate in the co-moving frame corresponds to the increased appearance of phase defects in the laboratory frame. A decrease of phase defects can be observed from $B = 0.05\,\mathrm{T}$ to $B = 0.08\,\mathrm{T}$ and vise versa an increase from $B = 0.08\,\mathrm{T}$ to $B=0.13\,\mathrm{T}$. This indicates a reduction of irregularity when the plasma enters the phase-locked regime and an increase when it leaves. Thus, puff splitting is important for the transition to turbulence also in magnetized plasma. 


To identify turbulent patches only values $\geq40\,\%$ of $\widetilde{\Phi_{\mathrm{fl}}}{}_{{\mathrm{,max}}}$ are considered. This threshold is arbitrary to a certain degree. By means of a flood-fill algorithm, cohesive structures are found. The time difference of its appearance and disappearance is defined as its lifetime. This procedure is illustrated in fig.~\ref{Fig_examples2}. Figure~\ref{Fig_examples2}~a is a corrected time window, as known from the column~2 of fig.~\ref{Fig_examples}. Black patches in fig.~\ref{Fig_examples2}~b, found by the algorithm are labeled with the numbers 1 to 6 and only the lifetimes of 3,4 and 6 are indicated by double headed arrows, although all six patches are taken into account for evaluation.
But we can also assign the patches to each other. Again by means of a flood-fill algorithm cohesive structures are found with values $\geq10\,\%$ of $\widetilde{\Phi_{\mathrm{fl}}}{}_{{\mathrm{,max}}}$. This way if patches ($\widetilde{\Phi_{\mathrm{fl}}}\geq0.4\,\widetilde{\Phi_{\mathrm{fl}}}{}_{{\mathrm{,max}}}$) are connected by a path ($\widetilde{\Phi_{\mathrm{fl}}}\geq0.1\,\widetilde{\Phi_{\mathrm{fl}}}{}_{{\mathrm{,max}}}$), the younger patch can be assumed to be a split of the already existing patch.
Their split times are defined to be the difference of their respective starting times. Fig.~\ref{Fig_examples2} c illustrates this process. Paths that remained grey do not yield any splitting times. The small path at the bottom right and left side does not contain an actual patch at all and patch number 5 is not connected.
However, structures 1 and 3 are connected as indicated in yellow, as well as 2, 4 and 6 as indicated in orange. Thus, this example will yield three splitting times, which are represented by blue double headed arrows.

Decay and splitting are thought to be memory-less stochastic processes. As such they are describable by an exponential distribution $\exp(-t/t_d)$. In this case, the mean lifetime and splitting time can be calculated from the decay times $t_d$ of the exponential distribution according to $\langle t\rangle=t_d$. Fig.~\ref{Fig_examples2} d shows the life- and splitting times packed into $20$ bins for a time window of $300\,\mathrm{ms}$. Finally an exponential fit (dashed/dotted lines) was utilized to determine the average life- and split times for a given magnetic field strength. 

\begin{figure}[tbh]
    \centering
    \includegraphics[width=0.49\textwidth]{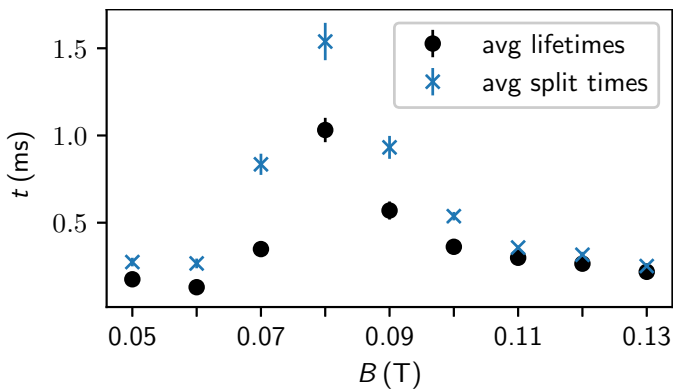}
	   \caption{Averaged lifetimes and splitting times in dependence of the magnetic field strength. Values for each magnetic field were calculated as outlined in fig.~\ref{Fig_examples2}.}
       \label{Fig_examples3} 
\end{figure}

Figure~\ref{Fig_examples3} shows the averaged lifetimes and splitting times in dependence of the magnetic field strength. In general, lifetimes and splitting times are of similar order of magnitude. In the quasi-linear regime ($B<0.07\,\mathrm{T}$), the splitting time is well above the life time, approximately a factor of two. By increasing the magnetic field, the first transition to the phased locked regime occurs around $0.08\pm 0.01\,\mathrm{T}$, which is characterized by a particularly long-lived quasi-coherent mode. However, also the splitting time is similarly enhanced. The splitting time is roughly a factor of two above the lifetime in the phase locked regime. As soon as the phase locked regime breaks down ($B<0.09\,\mathrm{T}$), we are in the weakly turbulent regime. In contrast to the pipe flow, the lifetime decreases with increasing control parameters. However, the splitting time also decreases and approaches the lifetime more and more with increasing control parameters. The turbulence approaches persistent turbulence as described in Ref.~\cite{AVILA2010}.  


In Summary, the dynamics at the transition in turbulence in magnetized plasmas was studied for analogies to pipe flows. In pipe flows, self-sustaining turbulence develops when the turbulent structures, called puffs there, divide on average more frequently before they decay \cite{AVILA2010}. Even though flows of neutral fluids through pipes and in cylindrical magnetized plasma appear very similar in geometry, there are very different instabilities that drive both systems into or sustain turbulence. Whether the concept can be transferred from pipe flows to magnetized plasmas is the objective of this study. The transition to turbulence was investigated in the linear plasma experiment PANTA \cite{YAMADA2008}. Signs of puff splitting, known from pipe-flows, were found in the experiment. Phase defects of waves in the laboratory frame were found to correspond to splitting of turbulent patches in the co-moving frame. The Reynolds number is not suitable as a control parameter in magnetized plasmas. As in previous work in helicon heated plasmas \cite{BURIN2005,MANZ2011a,THAKUR2014}, the magnetic field strength was chosen as the control parameter.
With respect to the magnetic field as a control parameter, three regimes are shown to be in agreement with the Landau-Ruelle-Takens scenario, the quasi-linear regime ($B\lesssim 0.07\,\mathrm{T}$), the phase locked regime ($0.07\lesssim B/\mathrm{T} \lesssim 0.09)$ and the weakly turbulent regime ($B\gtrsim 0.09\,\mathrm{T}$). In the quasi-linear and phase locked regimes, the splitting time is well above the lifetime. In the weakly turbulent regime, contrary to the pipe flow, the lifetime drops with the control parameter. However, the splitting time decreases even more, approaching the lifetime with increasing control parameter towards persistent turbulence.    

This work was partly supported by JSPS KAKENHI Grant Numbers 22H00120 and  21K13898. 


\end{document}